\documentstyle{article}

\begin{document}

\title {Models of Flavor with Discrete Symmetries}

\author{Alfredo Aranda~\footnote{aranda@bu.edu} \\
Department of Physics, Boston University.}
 
\maketitle

\begin{abstract}
In an attempt to understand the observed patterns of lepton and quark
masses, models invoking a flavor symmetry $G_f$, under which the Standard
Model generations are charged, have been proposed. One particularly
successful symmetry, U(2), has been extensively discussed in the
literature. The Yukawa matrices in models based on this symmetry reproduce 
the observed mass ratios in the lepton and quark sectors. The features of
the symmetry that determine the texture of the Yukawa matrices can be found
in other symmetries as well. We present a model based on a minimal, 
non-Abelian discrete symmetry that reproduces the Yukawa matrices associated
with U(2) theories of flavor. In addition to reproducing the mass and 
mixing angle relations obtained in such theories, the different representation
structure of our new horizontal symmetry allows for solutions to the solar and
atmospheric neutrino problems.
\end{abstract}

\paragraph{Introduction}

In this talk we discuss the possibility of using discrete symmetries to 
construct models of flavor. We start with the observation
that a U(2) symmetry has been
used~\cite{u2papers} to construct a successful model of 
quarks and charged leptons where
all the mass ratios and mixing angles are generated via two small
parameters associated with the breaking of the flavor symmetry.
It is interesting to ask whether there is a smaller symmetry that
can reproduce the results of the U(2) model and can also
be extended to incorporate the recent results on neutrino mixing.
This symmetry does indeed exist and it was discussed in \cite{tprime}.
There it was shown that using $T^{\prime} \times Z_3$ symmetry
one can construct a viable and minimal model of flavor. In the next
section we review the basic features of the U(2) model and 
point out the key ingredients that a symmetry must have to
generate the desired Yukawa textures. We then present the
necessary steps to construct a model using the new 
local discrete symmetry and a minimal model is outlined. Finally
some comments on possible implementations of $T^{\prime}$ 
beyond the minimal model are presented before concluding.

\paragraph{U(2) Model}

The flavor symmetry group is $G_f = $U(2). Quarks and leptons
of the first two generations
are assigned to a ${\bf 2}$ representation while the third
generation fields are singlets. The
assignment of the third generation as a singlet is motivated by 
the heaviness of the top quark. Putting the first two generations in 
a doublet yields degenerate scalar masses and thus the model is
safe from FCNC contributions.
The model contains three flavon fields: $\phi$ transforming as a doublet,
$A$ a singlet, and $S$ a triplet. When these flavons acquire vevs they
break the flavor symmetry. The breaking occurs in 
two steps, the first one is
generated by the vevs of $\phi$ and $S$. This happens in such a way that
a U(1) symmetry that rotates first generation fields by a phase is left
unbroken. This remaining U(1) is then broken down to nothing at a somewhat
lower scale by the vev of $A$. The result is a set of Yukawa textures
described by two parameters, $\epsilon$ which is related to the vevs of
$\phi$ and $S$, and $\epsilon^{\prime}$ related to the vev of $A$ ( and
therefore $\epsilon^{\prime} < \epsilon$). For a detailed description 
of the model see Refs.~\cite{u2papers}
The ingredients that are key in obtaining the U(2) Yukawa textures are:
the ${\bf 1}$, ${\bf 2}$, and ${\bf 3}$ representations of
U(2) are used in the model; the multiplication rule ${\bf 2} \otimes
{\bf 2}$ puts the vevs of $A$ in the right place and with the right 
sign; the existence of a U(1) subgroup that rotates first generation
fields by a phase. These are the key ingredients that a smaller symmetry
must contain in order to reproduce the successful textures of the U(2) 
model.

\paragraph{$T^{\prime}$}

The group $T^{\prime}$ is the smallest with ${\bf 1}$, ${\bf 2}$,
and ${\bf 3}$ dimensional representations with the desired 
multiplication rule.~\cite{tprime} Therefore we use $T^{\prime}$
to construct the minimal model. The representations are
${\bf 1}^{0}$, ${\bf 1}^{\pm}$,${\bf 2}^{0}$, ${\bf 2}^{\pm}$,
and ${\bf 3}$. The superscripts add modulo 3.
The remaining step is to determine 
whether or not it is possible to find a subgroup that allows one
to break the symmetry sequentially and generate the desired textures. 
$T^{\prime}$ has a $Z_3$ subgroup which can be used as the remaining
symmetry during the first breaking, namely, a symmetry that rotates
first generation fields by a phase. The two-dimensional representation
matrix of the element that generates this subgroup and that corresponds
to the desired rotation turns out to correspond to ${\bf 2}^{-}$. 
Unlike the U(2) model however, there is an additional
condition that must be satisfied which did not exist before. 
We argued that it would be interesting to find a smaller
symmetry, hence a discrete symmetry is desirable. Furthermore, we
are interested in the possibility of having a ``local'' discrete
symmetry. This is motivated by several arguments that global 
symmetries are violated by quantum gravitational effects.~\cite{col:1988}
If this is the case we need to make sure the model is anomaly free. 
This can be done by noting that $T^{\prime}$ is a subgroup of
SU(2) and thus can be embedded in it. If we do this, then the only
constraint on the model is that the matter fields fill out complete
SU(2) representations, which correspond to the ${\bf 2}^{0}$ and
${\bf 1}^{0}$ reps of $T^{\prime}$ (see \cite{tprime} for details).
This, together with the fact that the desired subgroup must rotate first
generations fields by a phase, leads us to extend the flavor symmetry
to $G_f = T^{\prime} \times Z_3$, this is the smallest group that
has the desired features. The two step breaking now can take place,
where the middle step symmetry is the diagonal $Z_3^D$ subgroup of
$G_f$. We assume that the $Z_3$ factor may be embedded in a U(1) gauge
symmetry whose anomalies are canceled by the Green-Schwarz 
mechanism~\cite{GS}.

\paragraph{A Model}
The three generations of matter fields are assigned to the representations
${\bf 2}{0-} \oplus {\bf 1}^{00}$ (the second triality corresponds to 
the $Z_3$ and also adds modulo 3). The Higgs fields $H_{U,D}$ transform
as singlets. The Yukawa mass matrices can now be obtained and we introduce
three flavons $A$, $\phi$, and $S$
with the representations ${\bf 1}^{0-}$, ${\bf 2}^{0+}$,
and ${\bf 3}^{-}$ respectively. Again, the vevs of $S$ and $\phi$ are
assumed to break $T^{\prime} \times Z_3$ down to $Z_3^D$ putting
entries of O($\epsilon$) in the Yukawa matrices, and then finally 
the vev of $A$ breaks the remaining $Z_3^D$ down to nothing yielding
entries of O($\epsilon^{\prime}$. These considerations yield the 
textures
\begin{eqnarray}\label{eq:textures}
Y_{U,D,L} \sim \left( \begin{array}{ccc}
0 & \epsilon^{\prime} & 0 \\
-\epsilon^{\prime} & \epsilon & \epsilon \\
0 & \epsilon & 1 \end{array} \right) \,\, ,
\end{eqnarray}
where O($1$) coefficients have been omitted. These are the same textures of
the U(2) model, as desired. As a note we mention that in order to 
differentiate between the up-type and down-type quarks it is possible to
embed both the U(2) model and the $T^{\prime} \times Z_3$ model into 
a GUT, for example an SU(5)~\cite{tprime}. Now the flavons may have
non-trivial transformation properties under the GUT symmetry and the
textures are accordingly modified. From now on the discussion will
concentrate on this ``GUT-model'' version.
Now that we have reproduced the U(2) model, neutrinos are introduced into
the model. Three generations of right-handed neutrinos are introduced 
with the assignment ${\bf 2}^{0-}\oplus {\bf 1}^{-+}$. This assignment
leads to Dirac and Majorana mass matrices that allow the introduction of
flavons that do not contribute at all to the charged fermion mass matrices.
Two such flavons are introduced transforming as ${\bf 2}^{+0}$ and yielding
the following Dirac and Majorana mass matrices:

\begin{eqnarray} \label{mlrmrrgut}\nonumber
M_{LR}  & \approx & \left( \begin{array}{ccc} 0 
& l_1 \epsilon' & l_5 r_2 \epsilon' \\ -l_1
\epsilon' & l_2 \epsilon^{2} & l_3 r_1 \epsilon \\ 0 &
l_4 \epsilon & 0 \end{array} \right) \langle H_U \rangle \,\,\, , \\
M_{RR}  & \approx & \left( \begin{array}{ccc} r_3 {\epsilon'}^{2} 
& r_4 \epsilon \epsilon' & r_2 \epsilon' \\ r_4
\epsilon \epsilon' & r_5 \epsilon^{2} & r_1 \epsilon \\ r_2 \epsilon' &
r_1 \epsilon & 0 \end{array} \right) \Lambda_R \,\,\, ,
\end{eqnarray}
where O($1$) coefficients have been introduced and $\Lambda_R$ is the
right-handed neutrino scale. Using the seesaw mechanism
 one obtains
the texture
\begin{eqnarray} \label{MLLGUT}
M_{LL}  \sim  \left( \begin{array}{ccc} (\epsilon'/\epsilon)^{2} & 
\epsilon{'}/\epsilon & \epsilon{'}/\epsilon 
\\ \epsilon{'}/\epsilon & 1 & 1 
\\ \epsilon{'}/\epsilon & 1 & 1
\end{array} \right) \frac{\langle H_U \rangle^2}{\Lambda_R} \,\,\, .
\end{eqnarray}
This texture leads naturally to large mixing between second and 
third generation neutrinos. The $1-2$ mixing is of 
O($\epsilon^{\prime}/\epsilon$), which can be accommodated to give the
bimaximal solution with the use of the O($1$) coefficients (in order
to determine the mixings accurately one computes the CKM matrix for
the lepton sector). In \cite{tprime} we presented a detailed  numerical
analysis of this model consisting of a fit to
the experimental data. This fit contained a renormalization group 
analysis and a $\xi^2$ minimization in order to prove that a set
of O($1$) coefficients could be found that reproduced the experimental
data. 

\paragraph{Alternative uses}
Here we comment on the possibility of using the group $T^{\prime}$ in 
different models of flavor. In particular, it can be used as a global
symmetry. In this case there is no need for an extra $Z_3$ and it is
possible to have a model that reproduces the U(2) model and accommodates
the solutions to the atmospheric and solar neutrino deficits~\cite{tprime}.
This is an interesting result when one notes that $T^{\prime}$ is a
subgroup of SU(2), which is known not to lead to a good theory of
flavor unless flavor universality is assumed. 
Another example in which $T^{\prime}$ can be used is to
consider the model based on the local $T^{\prime} \times Z_6$ symmetry
presented in \cite{tprime}. In this model it is not necessary to have
a GUT in order to explain the differences between the up- and down-type
quark sectors of the theory. Furthermore, this model also predicts
the ratio of $m_{t}/m_{b}$, which in the models described above it is 
put in by hand. This model also accommodates the neutrino results.

\paragraph{Conclusion}
Models based on $T^{\prime}$ flavor symmetry were discussed. 
In particular a minimal model with $G_f=T^{\prime} \times Z_3$ 
that reproduces the U(2) textures for fermion masses was reviewed. 
This model can also accommodate the results on neutrino oscillations. 
The main ingredients in the construction of the model were discussed.

\paragraph{Acknowledgments}
This work was done in collaboration with Christopher D. Carone and Richard F.
Lebed. The author is supported by the Department of Energy under
grant DE-FG02-91ER40676.

\bibliography{sample}

\begin{thebibliography}{99}
\bibitem{u2papers} R. Barbieri, G. Dvali, and L.J. Hall,  Phys.\ Lett.\ B {\bf 377}, 76 (1996);
R. Barbieri, L.J. Hall, and A. Romanino, Phys.\ Lett.\ B {\bf 401}, 47
(1997);
R. Barbieri, L.J. Hall, S. Raby and A. Romanino, Nucl.\ Phys.\ {\bf B493}, 
3 (1997).

\bibitem{tprime}A.~Aranda, C.~D.~Carone and R.~F.~Lebed,
arXiv:hep-ph/0010144;
A.~Aranda, C.~D.~Carone and R.~F.~Lebed,
Phys.\ Rev.\ D {\bf 62}, 016009 (2000)
[arXiv:hep-ph/0002044];A.~Aranda, C.~D.~Carone and R.~F.~Lebed,
Phys.\ Lett.\ B {\bf 474}, 170 (2000)
[arXiv:hep-ph/9910392].

\bibitem{col:1988} S.~R.~Coleman,
Nucl.\ Phys.\ B {\bf 310}, 643 (1988);
S.~B.~Giddings and A.~Strominger,
Nucl.\ Phys.\ B {\bf 307}, 854 (1988);
G.~Gilbert,
Nucl.\ Phys.\ B {\bf 328}, 159 (1989).

\bibitem{GS}
M. Green and J. Schwarz, Phys.\ Lett.\ B {\bf 149}, 117
(1984).

\end{thebibliography}

\end{document}